\documentclass{article}
\usepackage{graphicx}
\usepackage{amssymb}
\usepackage{txfonts}
\usepackage{hyperref}
\newcommand\rs[1]{_\mathrm{#1}}                 

\begin{document}

\noindent
{\LARGE\bf Approximation for radiation power of electrons
due to} 

\noindent
{\LARGE\bf inverse-Compton process
in the black-body photon field}

\vspace{0.3cm}
\noindent
{\bf O. Petruk} 

\vspace{0.2cm}
\noindent
{\it
   Institute for Applied Problems in Mechanics and Mathematics, Naukova St.\ 3-b,
   Lviv 79060, Ukraine\\
   Astronomical Observatory, National University, Kyryla and Methodia St.\ 8, Lviv 79008, Ukraine 
}
\vspace{0.5cm}

{\bf Abstract} {\small An approximation for the inverse-Compton radiation power of electrons in the isotropic black-body 
photon field is presented. 
The approximation allows one to calculate inverse-Compton emissivity as integral over 
the energies of incident electrons rather than over the field 
photon energies. Such an approach allows for accurate modeling of IC 
emission of electrons with energy spectra being different from power-law, 
in situation where the CPU resources are limited. 
High accuracy of this approximation allows one to use it in a wide range of conditions, 
from Thomson to extreme Klein-Nishina limits. 
The approach adopted results also in some new analytic expressions representing known results in 
the Thomson limit.}

\section{Introduction}

Since the first map of supernova remnant (SNR) RX J1713.7-3946 in very high-energy 
gamma-rays obtained by H.E.S.S. (Aharonian et al. \cite{RX1713-2004nature}), there are few more 
SNRs with spatial distributions of VHE $\gamma$-ray emission reported 
(see TeV Gamma-ray Source Catalog\footnote{Mori M., TeV Gamma-ray Source Catalog [available at 
     \url{http://www.icrr.u-tokyo.ac.jp/~morim/TeV-catalog/}]} 
     and reviews Rowell et al. \cite{Rowell-2005}, Funk \cite{Funk-2008}). 
Surface brightness distribution of $\gamma$-ray emission of astrophysical objects 
is an important possibility to test models of kinetics of astroparticles as well as 
dynamics of magnetic field and turbulence in astrophysical plasma. 
Thus, an actual and important task is the modeling of the respective surface 
brightness distribution. 

Inverse Compton (IC) electron-photon interactions is one of the most important processes 
of gamma-ray production in SNRs. Application of the exact formalism of IC emission 
to the numerical calculation of SNR images requires quite powerfull CPU resources. 
In such a situation, an accurate approximation may be of interest because it considerably 
reduces CPU time and also results in some new formulae in classical analysis 
(Sect. \ref{ICdelta} and \ref{ICThomson}). 

A common approach is to deal with IC emissivity for a given energy of 
the initial (monochromatic) photons (e.g. Jones \cite{jones-68}, Blumenthal \& Gould \cite{Blum-Gould-70}) with  
assumption of a given shape of electron spectrum (power-law as a common choice). 
The resulting IC photon spectrum is then given by the integration over the energy 
distribution of the field photons. 
Such an approach is essential for special cases of the field photon energy 
distributions. 

It is known however, that the black-body radiation field may be used for 
IC emission in many astrophysical objects. 
In particular, for SNRs under typical conditions, 
one may consider just black-body photons (either with single or with few different temperatures 
representing CMB/IR/optical radiation). 
In SNRs with no clear IR emission assosiated, 
the contribution from CMB photons dominates  
the role of infrared and optical photons (see discussion in Appendix in Lazendic et al. \cite{Lazendic-et-al-04}). 
The IR/optical photon fields may contribute typically 10\%-15\% of the IC flux in such SNRs 
(Gaisser et al. \cite{Gaisser-et-al-98}, Baring et al. \cite{Baring-et-al-99}). 

In this note, we present an approximation for IC emissivity 
which may be applied to IC emission originating from the black-body photon field with some 
temperature $T$. Since $T$ is a parameter in our approach, the approximation may be used for calculation 
of IC radiation from different photon fields (CMB, IR, optical). 
The target radiation field around some SNRs (e.g. around Galatic center) 
may not be black-body and/or the contribution from IR/optical photons may dominate over CMB there (Porter et al. \cite{Porter-et-al-06}, Hinton \& Aharonian \cite{Hinton-Aharonian-07}). 
In cases when different components of the target radiation field may be approximated 
by a superposition of multiple Planck distributions with different $T$, our approximation 
may be used in a similar fashion. Namely, the overall IC emission will be the  
weigthed sum of single approximations, each with different value of the temperature. 
In cases when the initial radiation field may not be approximately 
described by a sum of black-body distributions, our approximation is not applicable. 

Another assumption in the present paper is the isotropy of the electron and photon fields. 
A thorough treatment of anisotropic IC scattering from cosmic-ray electrons is done by 
Moskalenko \& Strong (\cite{Moskalenko-Strong-00}). 

Our approximation is given in terms 
of an energy of incident electrons rather than in commonly used terms of the field 
photon energy. Such an approach opens the possibility for accurate modeling of IC 
emission of electrons with energy spectra being different from power-law. 
For example, if we are interested in electron energies around maximum possible values. 
It is known that contribution from electrons accelerated by the shock to 
$E\rs{max}\sim 30-300$ TeV is important in interpretation of the H.E.S.S. 
observations of shell SNRs. 

\section{Approximation}
\subsection{Overview of known formulae}

The spectral ditribution of the volume emissivity of (isotropically distributed) 
electrons due to IC process is 
(Jones \cite{jones-68}, Blumenthal \& Gould \cite{Blum-Gould-70}, Schlickeiser \cite{Schlick-book})
\begin{equation}
 P\left(E\rs{\gamma}\right)=cE\rs{\gamma}\int d\gamma N(\gamma)\int d\epsilon 
 n\rs{ph}(\epsilon)\sigma\rs{KN}\left(E_{\gamma},\epsilon;\gamma\right)
 \label{PICdef}
\end{equation}
where $\gamma$ is the Lorenz factor of electron, $N(\gamma)$ is the spectral distribution of electrons, $\epsilon$ and $E\rs{\gamma}$ are the energies of photon before and after interaction respectively, 
$ n\rs{ph}(\epsilon)$ is the (isotropic) initial photon energy distribution, 
\begin{equation}
	\sigma\rs{KN}\left(E_{\gamma},\epsilon;\gamma\right)=	\frac{3\sigma\rs{T}}{4\epsilon \gamma^2}G\left(q,\eta\right)
\end{equation}
is the angle-integrated IC cross-section, $\sigma\rs{T}$ is the Thomson cross-section, 
\begin{equation}
	G\left(q,\eta\right)=	2q\ln q+(1+2q)(1-q)+2\eta q(1-q),
	\label{Gdef}
\end{equation}
\begin{equation}
 q=\frac{E\rs{\gamma}}{\Gamma(\gamma m\rs{e}c^2-E\rs{\gamma})}, \qquad
 \Gamma=\frac{4\epsilon \gamma}{m\rs{e}c^2}, \qquad
 \eta=\frac{\epsilon E\rs{\gamma}}{(m\rs{e}c^2)^2}.
 \label{ICdefs}
\end{equation} 
The last term in $G$ reflects the Klein-Nishina decline; recoil is embedded in the $q$ parameter. The KN effect is more important than recoil for $\Gamma\geq 1$ (e.g. Baring et al. \cite{Baring-et-al-99}). Kinematic requirements result in $(4\gamma^2)^{-1}\leq q\leq 1$ (Blumenthal \& Gould \cite{Blum-Gould-70}). Setting $q$ to its minimum and maximum values limits energies of upscattered photons: 
\begin{equation}
 E\rs{\gamma,\min}=\frac{\gamma m\rs{e}c^2\Gamma}{4\gamma^2+\Gamma},
\end{equation}
\begin{equation}
 E\rs{\gamma,\max}=\frac{\gamma m\rs{e}c^2\Gamma}{1+\Gamma}
 \label{Egammamax}
\end{equation}
that simplifies to
\begin{equation}
 E\rs{\gamma,\min}=\epsilon,\qquad 
 E\rs{\gamma,\max}=4\gamma^2 \epsilon
\end{equation}
in Thomson limit ($\Gamma\ll 1$) and to 
\begin{equation}
 E\rs{\gamma,\min}=\epsilon\ (\mathrm{if}\ \Gamma\ll 4\gamma^2),\qquad 
 E\rs{\gamma,\max}=\gamma m\rs{e}c^2
\end{equation}
in extreme Klein-Nishina limit ($\Gamma\gg 1$). The condition $q\leq 1$ sets the minimum Lorentz factor 
\begin{equation}
	\gamma\rs{\min}=\frac{E\rs{\gamma}}{2m\rs{e}c^2}
	\left[1+\left(1+\frac{(m\rs{e}c^2)^2}{\epsilon E\rs{\gamma}}\right)^{1/2}\right]
	\label{gammamin}
\end{equation}
electron should have in order to scattter photon with energy $\epsilon$ to energy $E\rs{\gamma}$. The function $\gamma\rs{\min}(E\rs{\gamma})$ may approximately be splitted in two parts 
\begin{equation}
	\gamma\rs{\min}	=\left\{
	\begin{array}{ll}
	E\rs{\gamma}^{1/2}/\left(2\epsilon^{1/2}\right), &\ \mathrm{for}\quad\eta\ll 1\\	
	E\rs{\gamma}/\left(m\rs{e}c^2\right), &\ \mathrm{for}\quad\eta\gg 1\\
	\end{array}\right.\ .
	\label{Egammamin}
\end{equation}
(The Klein-Nishina decline is negligible for $\eta\ll 1$, Eq.~\ref{Gdef}.)
The point where one could approximately switch from $\gamma\rs{\min}\propto E\rs{\gamma}^{1/2}$ to $\gamma\rs{\min}\propto E\rs{\gamma}$ is 
\begin{equation}
 E\rs{\gamma,*}=\frac{\left(m\rs{e}c^2\right)^2}{4\epsilon}.
\end{equation}

\subsubsection{Method of approximation}

In some astrophysical environments, the initial 
photon energy field may well be represented by the isotropic black-body radiation 
\begin{equation}
 n\rs{ph}(\epsilon)=\frac{1}{\pi^2 \hbar^3 c^3 }
 \frac{\epsilon^2}{\exp\left(\epsilon/ \epsilon\rs{c}\right)-1}
 \label{seedphotons}
\end{equation}
with $\epsilon\rs{c}=kT$. 

Let us re-write Eq.~(\ref{PICdef}) in the form
\begin{equation}
 P\left(E\rs{\gamma}\right)=\int d\gamma N(\gamma)p(\gamma,E\rs{\gamma})
 \label{ICemiss}
\end{equation}
where the spectral distribution of IC radiation power of a "single" electron
with Lorenz factor $\gamma$ is 
\begin{equation}
 p(\gamma,E\rs{\gamma})=
 \frac{3\sigma\rs{T}m\rs{e}^2c^2\epsilon\rs{c}}{4\pi^2 \hbar^3}
 \gamma^{-2}{\cal I}(\eta\rs{c},\eta\rs{o})
 =\frac{2e^4 \epsilon\rs{c}}{\pi \hbar^3c^2}
 \gamma^{-2}{\cal I}(\eta\rs{c},\eta\rs{o})
 \label{ICpower}
\end{equation}
with the function ${\cal I}(\eta\rs{c}(E\rs{\gamma}),\eta\rs{o}(\gamma,E\rs{\gamma}))$ 
\begin{equation}
 {\cal I}(\eta\rs{c},\eta\rs{o})=
 \int \frac{(\eta/\eta\rs{c})G(\eta\rs{o}/\eta,\eta)}{\exp\left(\eta/\eta\rs{c}\right)-1}d\eta,
 \label{calIIC}
\end{equation}
\begin{equation}
 \eta\rs{c}={\epsilon\rs{c}E\rs{\gamma}\over \left(m\rs{e}c^2\right)^2}, \qquad 
 \eta\rs{o}\equiv q\eta={E\rs{\gamma}^2\over 4\gamma m\rs{e}c^2(\gamma m\rs{e}c^2-E\rs{\gamma})}.
\end{equation}

Let us introduce 
\begin{equation}
	G_1\left(q\right)=	2q\ln q+(1+2q)(1-q),
\end{equation}	
\begin{equation}	
	G_2\left(q,\eta\rs{o}\right)=	2\eta\rs{o} (1-q).
\end{equation}
In the limit $\eta\rightarrow\infty$,  $G_1,\ G_2,\ G$ approach assymptotycally the values
\begin{equation}
	G\rs{1,as}=	1,\qquad
	G\rs{2,as}=	2\eta\rs{o},\qquad
  G\rs{as}=1+2\eta\rs{o}.
\end{equation}
The minimum value of $\eta$, namely $\eta\rs{\min}=\eta\rs{o}$, is given by the condition $G(\eta\rs{o}/\eta,\eta\rs{o})=0$. Relation $\eta\rs{\min}=\eta\rs{o}$ with definition of $\eta$, Eq.~(\ref{ICdefs}), yeild the formula for the minimum energy of the photon $\epsilon\rs{\min}$ which may be upscattered to the energy $E\rs{\gamma}$ by the electron with Lorentz factor $\gamma$: 
\begin{equation}
 \epsilon\rs{\min}=\frac{E\rs{\gamma}m\rs{e}c^2}{4\gamma\left(\gamma m\rs{e}c^2-E\rs{\gamma}\right)}.
\end{equation}

In the limit $\eta\rs{o}\ll\eta\rs{c}$, that is equivalent to the Thomson limit $\Gamma(\epsilon\rs{c})\ll 1$, the integral (\ref{calIIC}) may be found analytically 
\begin{equation}
 {\cal I}\rs{T}(\eta\rs{c},\eta\rs{o})= 
 \eta\rs{c}\int\limits_{0}^{\infty}\frac{\eta'd\eta'}{\exp(\eta')-1}=
 \frac{\pi^2\eta\rs{c}}{6}. 
 \label{app3}
\end{equation}

With decreasing of $\eta$, $G$ falls rather rapidly from $G\rs{as}$ to zero. 
Let $G(\eta\rs{o}/\eta,\eta)$ be approximated by Heavicide step function 
\begin{equation}
 G(\eta\rs{o}/\eta,\eta)\approx (1+2\eta\rs{o}) H(\eta-\eta\rs{o}).
\end{equation}
The integral ${\cal I}$ may then be approximately found as ${\cal I}(\eta\rs{c},\eta\rs{o})\approx 
 {\cal I}\rs{H}(\eta\rs{c},\eta\rs{o})$: 
\begin{equation}
\begin{array}{ll}
 &\displaystyle
 {\cal I}\rs{H}(\eta\rs{c},\eta\rs{o})
 =(1+2\eta\rs{o})
 \int\limits_{\eta\rs{o}}^{\infty} \frac{\eta/\eta\rs{c}}{\exp\left(\eta/\eta\rs{c}\right)-1}d\eta\\
 &\displaystyle \quad
 =(1+2\eta\rs{o})\eta\rs{c}\left(\frac{\pi^2}{6}+
  \mathrm{Li_2}(\exp(\eta\rs{o}/\eta\rs{c}))+\frac{\left(\eta\rs{o}/\eta\rs{c}\right)^2}{2}\right)
\end{array}  
  \label{app1}
\end{equation}
where the dilogarithm function 
\begin{equation} 
 \mathrm{Li_2}(x)=-\int\limits^{1}_{x} \frac{\ln(t)}{1-t}dt.
\end{equation}
An accurate approximation of (\ref{app1}) is
\begin{equation}
 {\cal I}\rs{H}(\eta\rs{c},\eta\rs{o})\approx 
 \frac{\pi^2}{6}\eta\rs{c}(1+2\eta\rs{o})\exp\left(-\frac{2\eta\rs{o}}{3\eta\rs{c}}\right).
  \label{app2}
\end{equation}
This expression restores ${\cal I}$ almost exactly for $\eta\rs{o}/\eta\rs{c}\lesssim10^{-2}$ and 
$\eta\rs{o}/\eta\rs{c}\gtrsim1$. 
However, ${\cal I}\rs{H}(\eta\rs{c},\eta\rs{o})$ overestimates ${\cal I}(\eta\rs{c},\eta\rs{o})$ up to 2 times for $10^{-2}\lesssim\eta\rs{o}/\eta\rs{c}\lesssim1$. This is because the deviation of $G$ from Heavicide step function is important in this range.  

The original integral ${\cal I}$ has an important property. Namely, it may be splitted as ${\cal I}={\cal I}_1+{\cal I}_2$ with  
\begin{equation}
 {\cal I}_1(\eta\rs{c},\eta\rs{o})=
 \int \frac{(\eta/\eta\rs{c})G_1(\eta\rs{o}/\eta)}{\exp\left(\eta/\eta\rs{c}\right)-1}d\eta
 = \eta\rs{c}\int \frac{xG_1(x\rs{o}/x)}{\exp\left(x\right)-1}dx, 
 \label{calIICsplit1}
\end{equation}
\begin{equation}
\begin{array}{ll}
 {\cal I}_2(\eta\rs{c},\eta\rs{o})
 &\displaystyle
 = \eta\rs{o}
 \int \frac{(\eta/\eta\rs{c})G_{2*}(\eta\rs{o}/\eta)}{\exp\left(\eta/\eta\rs{c}\right)-1}d\eta\\ \\
 &\displaystyle
 =\eta\rs{c}\eta\rs{o}\int \frac{xG_{2*}(x\rs{o}/x)}{\exp\left(x\right)-1}dx,
\end{array} 
 \label{calIICsplit2}
\end{equation}
where $x\rs{o}=\eta\rs{o}/\eta\rs{c}$ and $G_{2*}=G_2/\eta\rs{o}$. As one can see, these integrals may be scaled: 
\begin{equation}
 {\cal I}_1(a\eta\rs{c},a\eta\rs{o})=a{\cal I}_1(\eta\rs{c},\eta\rs{o}),
 \quad
 {\cal I}_2(a\eta\rs{c},a\eta\rs{o})=a^2{\cal I}_2(\eta\rs{c},\eta\rs{o}).
 \label{Iscale}
\end{equation}
The possibility to scale these summands (integral ${\cal I}$ may not be universally scaled), 
is an important property which allows us to obtain an analytic approximation. It is also the reason of 
accuracy of approximation over the wide range of parameters. 

   \begin{figure}
   \centering
   \includegraphics[width=8truecm]{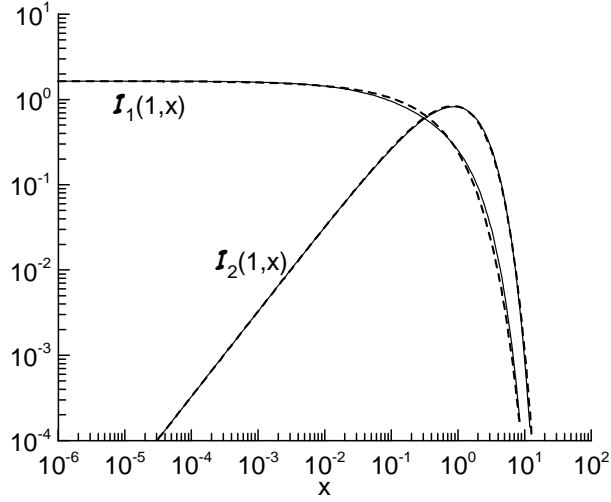}
      \caption{The accuracy of the approximation (\ref{calIappranyeta}) as sum ${\cal I}_1+{\cal I}_2$. 
      Integrals ${\cal I}_1(1,x)$, 
      ${\cal I}_2(1,x)$ (solid lines) are compared here with their 
      approximations (\ref{I1appr}) and (\ref{I2appr}) (dotted lines).
      } 
         \label{fig-a}
   \end{figure}

Really, using $a=\eta\rs{c}^{-1}$ in the scale laws (\ref{Iscale}), one have 
\begin{equation}
 {\cal I}_1(\eta\rs{c},\eta\rs{o})=\eta\rs{c}{\cal I}_1(1,\eta\rs{o}/\eta\rs{c}),
 \quad
 {\cal I}_2(\eta\rs{c},\eta\rs{o})=\eta\rs{c}^2{\cal I}_2(1,\eta\rs{o}/\eta\rs{c}).
\end{equation}
{\it This means that it is enough to check how accurate will be approximate expressions 
in approximation of just ${\cal I}_1(1,x)$ and ${\cal I}_2(1,x)$, and we will know how accurate will be these  approximations for any $\eta\rs{c}$ and $\eta\rs{o}$.} 

Let us correct a bit exponential part in each summand of (\ref{app2}) by introducing into exponent the second term of the form $c_1(\eta\rs{o}/\eta\rs{c})^{c_2}$ where $c_1$ and $c_2$ are constant. These terms make the summands in (\ref{app2}) to be more accurate in the representation of ${\cal I}_1$ and ${\cal I}_2$. It is important that the terms are also scaled in accordance to (\ref{Iscale}). 
With these corrections, we come to approximations 
\begin{equation} 
 {\cal I}_1\approx 
 \frac{\pi^2}{6}\eta\rs{c} 
 \exp\left[-\frac{2\eta\rs{o}}{3\eta\rs{c}}-\frac{5}{4}\left(\frac{\eta\rs{o}}{\eta\rs{c}}\right)^{1/2}\right],
 \label{I1appr}
\end{equation}
\begin{equation} 
 {\cal I}_2\approx 
 \frac{\pi^2}{3}\eta\rs{c} 
 \eta\rs{o}
 \exp\left[-\frac{2\eta\rs{o}}{3\eta\rs{c}}-\frac{5}{7}\left(\frac{\eta\rs{o}}{\eta\rs{c}}\right)^{0.7}\right].
 \label{I2appr}
\end{equation}
The values of $c_1$ and $c_2$ are obtained by fitting the exact ${\cal I}_1(1,x)$ and ${\cal I}_2(1,x)$. 
They are compared with approximation on 
Fig.~\ref{fig-a} which reveals good accuracy of obtained approximate formulae. 

The scaling property is a reason, why the sum of approximations (\ref{I1appr}) and (\ref{I2appr}) 
\begin{equation}
\begin{array}{ll}
 {\cal I}(\eta\rs{c},\eta\rs{o})
 &\displaystyle 
 \approx \frac{\pi^2}{6}\eta\rs{c} \left(
 \exp\left[-\frac{5}{4}\left(\frac{\eta\rs{o}}{\eta\rs{c}}\right)^{1/2}\right]
 \right.\\ \\
 &\displaystyle\left.
 +2\eta\rs{o}
 \exp\left[-\frac{5}{7}\left(\frac{\eta\rs{o}}{\eta\rs{c}}\right)^{0.7}\right]
 \right)
 \exp\left[-\frac{2\eta\rs{o}}{3\eta\rs{c}}\right]
\end{array} 
 \label{calIappranyeta}
\end{equation}
is accurate to represent ${\cal I}$ in any regime, from Thomson to extreme Klein-Nishina.

\begin{figure}
   \centering
   \includegraphics[height=6.5truecm]{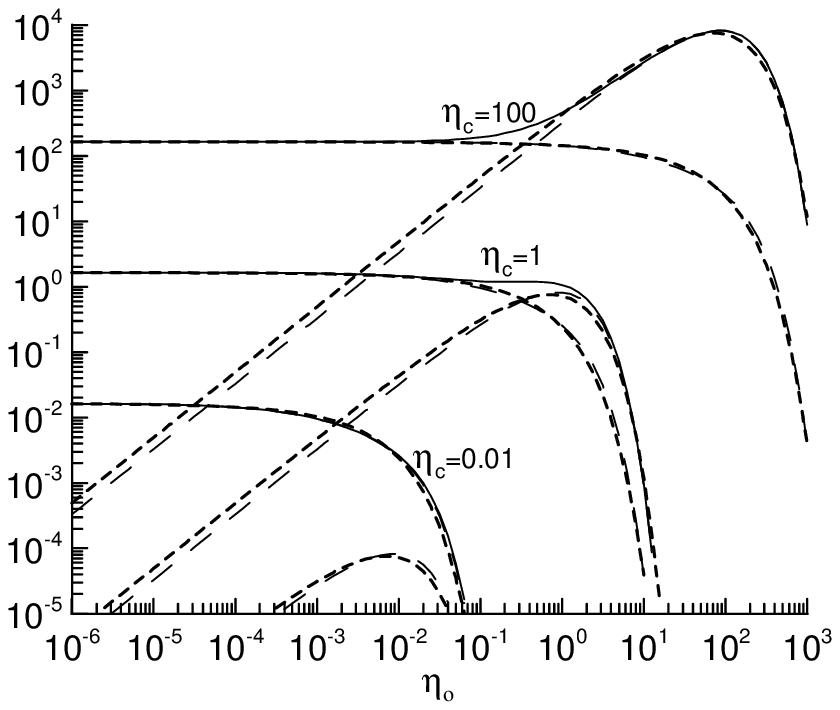}
      \caption{Accuracy of approximation (\ref{calIappr}). 
      Integrals ${\cal I}$ (\ref{calIIC}) (solid lines), 
      ${\cal I}_1$ and ${\cal I}_2$ (\ref{calIICsplit1}), (\ref{calIICsplit2}) 
      (dashed lines) and respective summands of approximation (\ref{calIappr}) 
      (dotted lines) versus $\eta\rs{o}$ for a number of $\eta\rs{c}=0.01,1,100$ 
      (from below).
      } 
         \label{fig-b}
   \end{figure}

\begin{figure}
   \centering
   \includegraphics[height=6.5truecm]{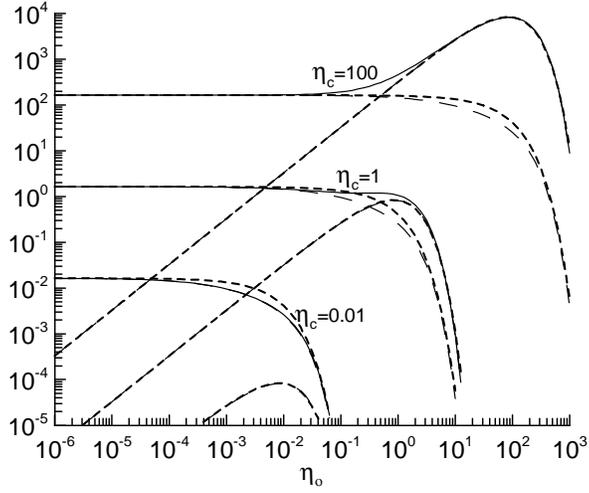}
      \caption{Accuracy of approximation (\ref{calIappreta100}). 
      Integrals ${\cal I}$ (\ref{calIIC}) (solid lines), 
      ${\cal I}_1$ and ${\cal I}_2$ (\ref{calIICsplit1}), (\ref{calIICsplit2}) 
      (dashed lines) and respective summands of approximation 
      (\ref{calIappreta100}) (dotted lines) versus $\eta\rs{o}$ 
      for a number of $\eta\rs{c}=0.01,1,100$ (from below).
      } 
         \label{fig-c}
   \end{figure}

Eq.~(\ref{calIappranyeta}) is good for any $\eta\rs{c}$. We may suggest two a bit simple approximations for different ranges of $\eta\rs{c}$. 
If one are interested mostly in $\eta\rs{c}\lesssim 100$ (the case of IC emission of electrons accelerated by the forward shock in SNRs) then one can use an expression
\begin{equation}
 {\cal I}(\eta\rs{c},\eta\rs{o})\approx 
 \frac{\pi^2}{6}\eta\rs{c} \left(1+3\eta\rs{o}\right)
 \exp\left[-\frac{2\eta\rs{o}}{3\eta\rs{c}}-\frac{5}{4}\left(\frac{\eta\rs{o}}{\eta\rs{c}}\right)^{1/2}\right].
 \label{calIappr}
\end{equation}
Fig.~\ref{fig-b} shows, that the second summand here ($\propto 3\eta\rs{o}$) overestimates ${\cal I}_2$ in its power-law part. Nevertheless, this error is negligible for $\eta\rs{c}\leq 100$ (Fig.~\ref{fig-b}). If $\eta\rs{c}\gtrsim 10$ is of interest, then one may neglect accuracy in exponential part of approximation of ${\cal I}_1$ and use the approximation (Fig.~\ref{fig-c})
\begin{equation}
 {\cal I}(\eta\rs{c},\eta\rs{o})\approx 
 \frac{\pi^2}{6}\eta\rs{c} \left(1+2\eta\rs{o}\right)
 \exp\left[-\frac{2\eta\rs{o}}{3\eta\rs{c}}-\frac{5}{7}\left(\frac{\eta\rs{o}}{\eta\rs{c}}\right)^{0.7}\right].
 \label{calIappreta100}
\end{equation}

Fig.~\ref{fig-d} shows the accuracy of Eq.(\ref{calIappr}) in approximation of emission power (\ref{ICpower}) for electrons with energies $10^{12}\div10^{16}$ eV.


\subsection{'Delta-function' approximation}
\label{ICdelta}

\begin{figure}
   \centering
   \includegraphics[height=6.5truecm]{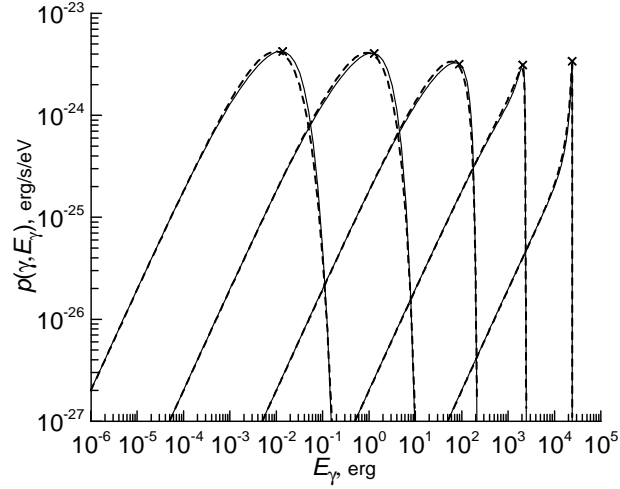}
      \caption{The spectrum $p(E\rs{\gamma})$ (\ref{ICpower}) 
      calculated with integral ${\cal I}$ (solid lines) and with its 
      approximation (\ref{calIappr}) (dashed lines) 
      for Lorentz factors 
      $\gamma=3\cdot (10^6,10^7,10^8,10^9,10^{10})$ (from the left) that 
      correspond to electron energies
      $E=1.5\cdot(10^{12},10^{13},10^{14},10^{15},10^{16})\ \mathrm{eV}$.
      Crosses correspond to position of $E\rs{\gamma,\max}(\gamma,\epsilon\rs{c})$.
      } 
         \label{fig-d}
   \end{figure}

Fig.~\ref{fig-d} demonstrates that `single' electron with Lorentz factor $\gamma$ -- 
being scattered by 
all black-body photons -- emits most of its IC radiation 
at photons with some characteristic energy $E\rs{\gamma m}$. 

One can introduce a 'delta-function approximation', in addition to 
the classical `monochromatic approximation' where electron is scattered by the photons with a 
fixed energy $\epsilon\rs{o}$ (e.g. Schlickeiser \cite{Schlick-book}). 
Namely, one can assume that 'single' electron (scattered by 
all black-body photons) emits all of its IC power at photons with 
$E\rs{\gamma m}$: 
\begin{equation} 
 p(\gamma,E\rs{\gamma})\approx p\rs{m}(\gamma)\delta(E\rs{\gamma}-E\rs{\gamma m})
 \label{deltaapprox}
\end{equation}
where
\begin{equation} 
 p\rs{m}(\gamma)=\int\limits_{0}^{\infty}p(\gamma,E\rs{\gamma}) dE\rs{\gamma}.
\end{equation}
There are well known approximations for the total IC energy loss $p\rs{m}$ of electron 
in the Thomson (see (\ref{totalElossT}) below) and extreme Klein-Nishina limits 
(e.g. Sect.~4.2.3 in Schlickeiser \cite{Schlick-book}). 

Our numerical calculations show (Fig.~\ref{fig-d}) that 
$E\rs{\gamma m}$ may well be approximated by 
\begin{equation} 
 E\rs{\gamma m}(\gamma)\approx E\rs{\gamma,\max}(\gamma,\epsilon\rs{c})
\end{equation}
where $E\rs{\gamma,\max}(\gamma,\epsilon)$ is given by (\ref{Egammamax}). 
In the Thomson limit, this is 
\begin{equation} 
 E\rs{\gamma m}(\gamma)\approx 4\epsilon\rs{c}\gamma^2.
\end{equation}
Note, that in the classical `monochromatic approximation', the average $\left\langle E\rs{\gamma}\right\rangle=(4/3)\epsilon\rs{o}\gamma^2$ is used as an estimator 
for the energy of `monochromatic' IC photons emitted by electron. 

\subsection{Thomson limit}
\label{ICThomson}

The use of (\ref{app3}) in (\ref{ICpower}) allows us to write down the expression for 
IC emissivity in the Thomson limit $\eta\rs{o}\ll\eta\rs{c}$. 
The spectral distribution of IC radiation power of a "single" electron with energy $E=\gamma m\rs{e} c^2$ is 
\begin{equation}
 p\rs{T}(\gamma,E\rs{\gamma})=
 \frac{\sigma\rs{T}  \epsilon\rs{c}^2}{8\hbar^3c^2} \frac{E\rs{\gamma}}{\gamma^{2}}
 =\frac{\pi e^4 \epsilon\rs{c}^2}{3\hbar^3c^2} \frac{E\rs{\gamma}}{E^2}, 
 \qquad E\rs{\gamma}\leq E\rs{\gamma,lim}
 \label{TomshICemis}
\end{equation}
where $E\rs{\gamma,lim}$ is a characteristic maximum energy defined below. 
This expression represents integration over all possible energies $\epsilon$ of the seed black-body photons. 

The power $p\rs{T}(\gamma,E\rs{\gamma})$ is increasing function of $E\rs{\gamma}$, 
while $p(\gamma,E\rs{\gamma})$ decreases rather rapidly after the maximum (Fig.~\ref{fig-d}). 
Let's define an energy $E\rs{\gamma,lim}$ by the condition 
\begin{equation}
 \int\limits_{0}^{E\rs{\gamma,lim}}p\rs{T}(\gamma,E\rs{\gamma})dE\rs{\gamma}=p\rs{mT}(\gamma)
 \label{ICdefElim}
\end{equation}
where 
\begin{equation}
 p\rs{mT}(\gamma)=(4/3)c\sigma\rs{T}\omega \gamma^2
 \label{totalElossT}
\end{equation}
is the total energy loss of electron due to IC in the Thomson limit, 
$\omega=\int \epsilon n\rs{ph}(\epsilon)d\epsilon$ is the energy density of all initial photons. 
The definition (\ref{ICdefElim}) results in 
\begin{equation}
 E\rs{\gamma,lim}=\frac{4}{\pi}\epsilon\rs{c}\gamma^2
 \left[\frac{2}{3}
 \int\limits_{0}^{\infty}\frac{z^{3} dz}{\exp(z)-1}
 \right]^{1/2}
 =2.65\epsilon\rs{c}\gamma^2.
\end{equation}
Note, that $E\rs{\gamma,lim}$ differs just a bit from $E\rs{\gamma,\max}(\epsilon\rs{c})=4\epsilon\rs{c}\gamma^2$. 

The volume emissivity of electrons distributed as
\begin{equation}
 N(\gamma)=N\rs{o}\gamma^{-s}\quad \mathrm{for}\quad \gamma\rs{\min}<\gamma<\gamma\rs{max},
 \quad \gamma\rs{\min}\ll\gamma\rs{\max}
 \label{Nepower}
\end{equation}
is 
\begin{equation}
 P\left(E\rs{\gamma}\right)=
 \frac{\sigma\rs{T}\epsilon\rs{c}^{2}}{8\hbar^3c^2(s+1)}N\rs{o}
 E\rs{\gamma}\gamma\rs{\min}^{-(s+1)}.
 \label{ICemissThomson}
\end{equation}
For $E\rs{\gamma}\lesssim E\rs{\gamma,*}$, the minimum Lorentz factor is $\gamma\rs{\min}=E\rs{\gamma}^{1/2}/(2\epsilon\rs{*}^{1/2})$, Eq.~(\ref{Egammamin}), and we come to 
approximation of (\ref{ICemiss}) in the Thomson limit
\begin{equation}
 P\left(E\rs{\gamma}\right)=
 \frac{2^{s-2}\sigma\rs{T}\epsilon\rs{c}^{2}\epsilon\rs{*}^{(s+1)/2}}{\hbar^3c^2(s+1)}N\rs{o}
 E\rs{\gamma}^{-(s-1)/2},
 \label{Thomsemiss}
\end{equation}
with the known slope $P\left(E\rs{\gamma}\right)\propto E\rs{\gamma}^{-(s-1)/2}$. 
The value of $\epsilon\rs{*}$ 
may be fixed by comparison of (\ref{Thomsemiss}) with e.g. expression (4.2.17) in Schlickeiser (\cite{Schlick-book}).  
Namely, $\epsilon\rs{*}={\cal A}(s)\epsilon\rs{c}$ with  
\begin{equation}
 {\cal A}(s)=\left[
 \frac{12}{\pi^2}\frac{(s^2+4s+11)}{(s+5)(s+3)^2}
 \int\limits_{0}^{\infty}\frac{z^{(s+3)/2} dz}{\exp(z)-1}
 \right]^{2/(s+1)}. 
\end{equation}
Numerically, ${\cal A}(1.8)=0.665$, ${\cal A}(2)=0.710$, ${\cal A}(2.2)=0.755$. 

\section{Conclusions}

Numerical evaluation of the spatial distribution of the IC emission in SNRs requires essential computational resources because the IC volume emissivity (\ref{PICdef}) consists in two enclosed integrations (over initial photon and electron energies). We developed the approximation (\ref{calIappranyeta}) of the spectral distribution of the IC emission power $p(E\rs{\gamma})$ of electrons with Lorentz factor $\gamma$ which are interacting with the isotropical black-body photon field. Namely, Eq.~(\ref{calIappranyeta}) restores known results (Blumenthal \& Gould \cite{Blum-Gould-70}) with high enough accuracy in any regime, from Thomson to extreme Klein-Nishina limits. It may be used for different astrophysical objects. For $kTE\rs{\gamma}\lesssim 100\left(m\rs{e}c^2\right)^{2}$, i.e. 
in case of IC $\gamma$-ray emission from electrons accelerated by the forward shock of SNRs, 
it is suitable to use a bit simple approximation (\ref{calIappr}). In the Thomson limit, our approach results in a simple expression (\ref{TomshICemis}).

Our approximation may be used in situation when the initial radiation field may be approximated with 
the Planck function with some temperature $T$ or when it may be represented by a superposition of the 
black-body distributions with different $T$. In addition, it assumes isotropy of electron and photon fields. 

The approximation is given in terms of an energy of incident electrons rather than in commonly used terms 
of the field photon energy (there are known approximation for the latter approach, see e.g. Schlickeiser \cite{Schlick-book}). 
Therefore, our approximation may be useful for analysis of the role of the electron spectrum 
with shapes different from power-law. 

The main idea behind our approach is the possibility to split initial integral into two parts which, contrary to the original integral, may be scaled. This scaling is the reason of accuracy of the approximation over the wide range of parameters, from Thomson to extreme Klein-Nishina regime. 

There is well known `monochromatic approximation' where electron is scattered by the monochromatic photons with energy $\epsilon\rs{o}$ (e.g. Schlickeiser \cite{Schlick-book}). Fig.~\ref{fig-d} shows that the spectral distribution of the radiation power of `single' electron with Lorentz factor $\gamma$ scattered by photons distributed with the Planck function is peaked at some energy $E\rs{\gamma m}(\gamma)$. This allows us to introduce -- similarly to the case of synchrotron emission -- the `delta-function' approximation for IC emission. In this approximation, Eq.~(\ref{deltaapprox}), all radiated energy of electron is assumed to be at $E\rs{\gamma m}(\gamma)$. 
In the Thomson limit, $E\rs{\gamma m}(\gamma)\approx 4\epsilon\rs{c}\gamma^2$ where $\epsilon\rs{c}=kT$. 
In the classical `monochromatic approximation', the average $\left\langle E\rs{\gamma}\right\rangle=(4/3)\epsilon\rs{o}\gamma^2$ is used as an estimator 
for the energy of emitted IC photons. 

Our approach results in some new expressions which represent known results. Namely, Eq.~(\ref{TomshICemis}) 
yeilds the spectral distribution of IC radiation power of a "single" electron and 
Eq.~(\ref{Thomsemiss}) represents the spectrum of IC emission from the power-law spectrum of electrons, 
in the Thomson limit. 
These expression account for integration over all possible energies of the seed black-body photons. 
They can be derived thanks to the possibility of analytical integration of ${\cal I}$ in the Thomson regime, Eq.~(\ref{app3}).



\end{document}